\documentclass[longauth,traditabstract,letter]{aa}
\usepackage{amsmath}
\usepackage{graphicx}
\usepackage{txfonts}

\begin{document}

\title{Dust in the bright supernova remnant N49 in the LMC
\thanks{Herschel is an ESA space observatory with science 
instruments provided by European-led Principal Investigator 
consortia and with important participation from NASA.}}

\institute{
Space Telescope Science Institute, 3700 San Martin Drive, Baltimore, 
MD 21218, USA\\
\and
School of Physical \& Geographical Sciences, Lennard-Jones 
Laboratories, Keele University, Staffordshire ST5 5BG, UK\\
\and
Department of Physics and Astronomy, University 
College London, Gower Street, London WC1E 6BT  \\
\and 
Spitzer Science Center, California Institute of Technology, 
MS 220-6, Pasadena, CA  91125 \\
\and
Stratospheric Observatory for Infrared Astronomy, 
Universities Space Research Association, Mail Stop 211-3, Moffett Field, CA 94035\\
\and
CEA, Laboratoire AIM, Irfu/SAp, Orme des Merisiers, 
F-91191 Gif-sur-Yvette, France\\
\and
Steward Observatory, University of Arizona, 933 
North Cherry Ave., Tucson, AZ 85721, USA\\
\and Jodrell Bank Centre for Astrophysics, Alan Turing Building, School
of Physics \& Astronomy, University of Manchester, 
Oxford Road, Manchester M13 9PL, United Kingdom  \\
\and  Department of Astronomy, Cornell University, Ithaca, NY 14853, USA \\
}

\author{
M. Otsuka\fnmsep\inst{1},
J. Th. van Loon\inst{2},
K. S. Long\inst{1},
M. Meixner\inst{1},
M. Matsuura\inst{3},
W.T. Reach\inst{4,5},
J. Roman-Duval\inst{1},
K. Gordon\inst{1},
M. Sauvage\inst{6},
S. Hony\inst{6},
K. Misselt\inst{7},
C. Engelbracht\inst{7},
P. Panuzzo\inst{6},
K. Okumura\inst{6},
 P.~M.~Woods\inst{8},
 F. Kemper\inst{8},
 \and  G. Sloan \inst{9}
}

\offprints{otsuka@stsci.edu}

\date{Received / Accepted}
\authorrunning{Otsuka et al.}
\titlerunning{Dust in the LMC SNR N49}

\abstract{
We investigate the dust associated with the supernova remnant (SNR) N49 
in the Large Magellanic Cloud (LMC) as observed with the {\it Herschel} 
Space Observatory. N49 is unusually bright because of an
interaction with a molecular cloud along its eastern edge. 
We have used PACS and SPIRE to measure
the far IR flux densities of the entire SNR and of a 
bright region on the eastern edge of the SNR where the 
SNR shock is encountering the molecular cloud. 
Using these fluxes supplemented with archival data 
at shorter wavelengths, we estimate the
dust mass associated with N49 to be about 10 $M_{\odot}$. 
 The bulk of the dust in our simple two-component model 
has a temperature of 20-30 K, similar to that of nearby 
molecular clouds. Unfortunately, as a result of the limited 
angular resolution of {\it Herschel} at the wavelengths sampled with 
SPIRE, the uncertainties are fairly large.  Assuming this estimate 
of the dust mass associated with the SNR is approximately correct, 
it is probable that most of the dust in the SNR arises from 
regions where the shock speed is too low to produce significant 
X-ray emission.
The total amount of warm 50-60 K dust is $\sim$0.1 
or 0.4 $M_{\odot}$, depending on whether the dust is modeled 
in terms of carbonaceous or silicate grains. This provides 
a firm lower limit to the amount of shock heated dust in N49.  
}

\maketitle

\section{Introduction}
N49 is a bright X-ray and optical supernova remnant (SNR) in the Large
Magellanic Cloud (Long et al. 1981), associated with 
a soft gamma ray repeater (Cline et al 1982) and 
unresolved X-ray point source (Rothschild et al. 1994). 
It has an asymmetric surface-brightness 
distribution at essentially all wavelengths, which is due to an interaction with a molecular 
cloud that is located on the southeast limb of the SNR (Vancura et
al. 1992; Banas et al 1997). Based on a simple Sedov model for the SNR,
Hughes et al. (1998) concluded that the swept-up mass was the order of
200 $M_{\odot}$. Park et al. (2003) recently conducted a detailed study of its X-ray properties 
using {\it Chandra}, confirming that the SNR is relatively young, on the
order of 6\,000 yrs.  They find that although most of the emission is
dominated by interstellar material (ISM), the {\it Chandra} observations appear 
to show ``explosion fragments'' on the SW side of the SNR in X-rays. 
According to Bilikova et al. (2007), the X-ray temperature
($\sim$1.2$\times$10$^{7}$ K) implies a shock velocity of 920 km
s$^{-1}$, while the optical filaments which
reflect shocks traversing denser gas have typical line-of-sight
velocities of 250 km s$^{-1}$. Hill et al. (1995) estimated the  
mass of the progenitor of N49 to be about 20 $M_{\odot}$ based on the association 
with the 10 Myr old association LH 53.
A 1720 MHz OH maser emission comes from a region toward the southwest.
N49 is almost certainly the result of a Type II explosion.

Apparently first detected at mid-IR wavelengths with {\it IRAS} (Graham et
al.\ 1987), the first mid-IR images of the SNR were obtained with {\it
Spitzer} (Williams, R. et al. 2006), where N49 is visible not
only in the MIPS 24 and 70 $\mu$m bands, but also in the IRAC 3.6, 4.5, 5.6
and 8 $\mu$m bands. IRS spectra of the eastern limb show strong line emission from 
O, Ne, Ar, Si, S, and Fe and H$_{2}$ in the 5-37 $\mu$m band 
(Williams, R. et al 2006). While lines and dust emission compete 
at the short wavelengths, van Loon et al (2010) showed with the low 
resolution MIPS spectra of the bright blob on the eastern limb that dust 
is the dominant source of emission at 70 $\mu$m, and
estimated a dust mass for N49 of $\sim$0.2 $M_{\odot}$.

Dust in SNRs can arise either from the SN itself or from the ISM.  
Ejecta dust is thought to play a key role as a coolant in the 
formation of high-redshift galaxies (see, e.g. Morgan \& Edwards 2003).  
However, given the amount of swept-up material in N49, most of 
the dust in N49 is expected to be of interstellar origin.

Here we describe observations of N49 obtained with the {\it Herschel} Space
Observatory. 
Our primary purpose is to explore the properties of dust in N49, and to
obtain a better estimate of the dust mass than was heretofore possible 
by probing the longer wavelength bands, where larger dust grains are 
expected to be more apparent. Combining {\it Herschel} data and data obtained 
from other observatories, we construct spectral energy distributions 
(SEDs) of the entire SNR and of a bright blob on the eastern limb, 
where the SNR is interacting with a molecular cloud.
(see, e.g. van Loon et al. 2010).  We use a simple two-temperature model
of the SED to estimate dust masses and temperatures.

\section{Observations and data reduction}

\begin{figure*}[t]
\centering
\includegraphics[width=14.0cm]{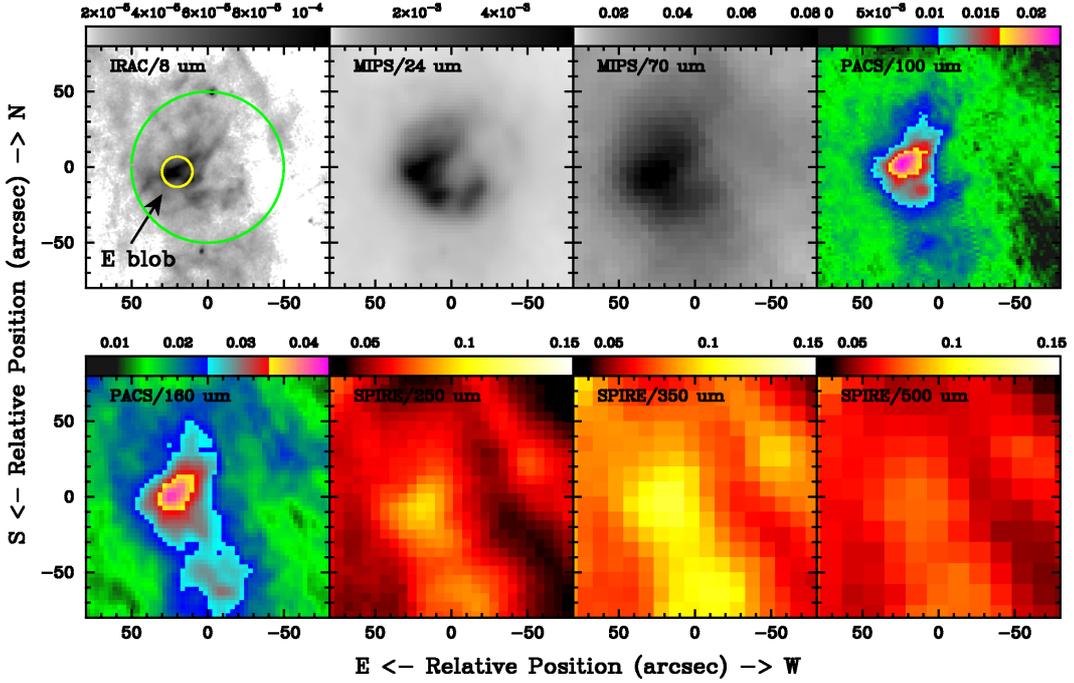}
\caption{Selected images of the SNR N49. The intensity is in Jy per
 pixel. The location of the eastern blob (E blob) is indicated by 
the arrow. The radii of the green and yellow
 circles are $50''$ and $10''$ when adopting aperture photometry for the
 whole SNR and the E blob.
\label{images}
}
\end{figure*}

N49 was observed with {\it Herschel} as part of the science demonstration program associated
with the HERITAGE program ( PI: M.Meixner). The observations and the
data reduction 
process for  the PACS green and red bands (100/160 $\mu$m) 
and SPIRE all bands (250/350/500 $\mu$m) are described by Meixner,  et al. 
(2010).

To complement the {\it Herschel} data, we used archival {\it Spitzer}/IRS low- and high-resolution 
spectra (5-40 $\mu$m; PI: Gehrz) of  the eastern blob (E blob, hereafter), 
{\it Spitzer}/IRAC 3.6/4.5/5.8/8.0 $\mu$m and MIPS 24/70 $\mu$m images (Meixner et al. 2006), 
and the published 3/7/11/15/24 $\mu$m fluxes for whole SNR (Seok et al. 2008) 
taken by the {\it AKARI}/IRC  (Murakami et al. 2007; Onaka et
al. 2007). We indicate the position of the E blob in Fig 1. 
We rescaled the flux density of the IRS spectrum by a factor of 1.25 in order to match the total flux between 
20.34 and 29.46 $\mu$m of the IRS spectrum and the MIPS 24 $\mu$m flux. We used the reductions of the 
IRS and the MIPS spectra (52-93 $\mu$m) of the E blob made by Kemper et al.\ (2010) as part the SAGE-Spec project. 
The MIPS spectrum has been described in detail by  van Loon et al.\
(2010).

Selected {\it Spitzer} and {\it Herschel} images are presented in Fig 1. Although the angular resolution
of PACS at 100 and 160 $\mu$m  is comparable to that of MIPS 24 and 70 $\mu$m, the resolution
of SPIRE is such that it becomes increasingly difficult to distinguish the SNR from the surrounding
molecular clouds.  The intensity of the E blob becomes comparable to
the southern molecular clouds at longer wavelengths. 
There is no obvious difference however in the morphology of the SNR
as a function of wavelength, although at long wavelengths, where the
resolution is poorer, it becomes increasingly difficult to pick out the shell of the SNR.  The
general interpretation that the SNR is interacting with a molecular
cloud on the SE is confirmed by the {\it Herschel} images.

To estimate flux densities, we performed aperture photometry.  We used 
50$''$ and 10$''$ radius 
regions to define the entire SNR and the eastern blob, respectively (see
Fig 1).  We
subtracted background from an annulus centered on the SNR with 
inner and outer radii of 110$''$ and 130$''$. In the IRAC 8 $\mu$m image 
of Fig 1, the two regions are indicated as green 
(50$''$ radius) and yellow (10$''$) circles. 
We performed aperture corrections using observed or theoretical 
point-spread functions. 
The measured fluxes are listed in Table 1. 
The absolute flux calibration is 2 and 5 $\%$ for MIPS 24 and 70 $\mu$m
bands and is 20 $\%$ for PACS (Poglitsch et al. 2010) and 15 $\%$ for 
SPIRE (Swinyard, Ade, \& Baluteau 2010), respectively. We adopted these values as
the flux density uncertainty for these bands. We did not include 
uncertainties associated with the subtraction of background, but 
these are at least as large in the SPIRE bands as the uncertainty 
in the flux calibration. For the other bands, the
uncertainty in Table 1 corresponds to the standard deviation of the
background.

The SEDs are presented in Fig 2 for the entire SNR and for the E-blob
alone.  There are no obvious differences aside from normalization in
the two spectra. Williams, R. et al.\ (2006), based on an early reduction 
of the IRS spectrum, argued that the emission lines rather than the 
dust continuum dominate the flux in the MIPS 24 $\mu$m band.  
However, our improved reduction of the  IRS spectrum
shows an obvious dust continuum.  Our estimate based on the full 9.7 $\mu
$m width of the MIPS 24 $\mu$m band is that emission lines 
contribute about 38 \% of the flux.  The fluxes plotted
at 24 $\mu$m in Fig 2 represent the continuum flux, assuming 
that the line contribution is the same for the entire
SNR as for the region covered by the IRS spectrum.  At wavelengths 
of $\lambda\sim$70 $\mu$m and longer, the radiation from dust 
grains in the SNR is the main luminosity source. van Loon et al.\ (2010) 
found dominant [O~{\sc i}] 63 $\mu$m line in the MIPS 70 $\mu$m band to contribute
only 11\% of the total emission in that band. Furthermore, based on this,  
the expected [O\,{\sc i}] 146 $\mu$m  ($^{3}P_{1}$-$^{3}P_{0}$)
flux is 3.4-7.2$\times$10$^{-13}$ erg s$^{-1}$ cm$^{-2}$ when assuming the electron
temperature to be 10$^{4}$ K and a density 10-10$^{4}$ cm$^{-3}$. The
contribution of [O\,{\sc i}] 146 $\mu$m in the PACS 160 $\mu$m band is
$\sim$7$\%$.

\begin{table}
\centering
\caption{Flux density of N49. 
(In the last column) 
(1): {\it AKARI} (Seok et al. 2008) (2): {\it Spitzer} (present work); 
(3): {\it Herschel} (present work).\label{flux}}
\begin{tabular}{rccc}
\hline
$\lambda$ &\multicolumn{2}{c}{$F_{\nu}$(Jy)}&\\
\cline{2-3}
($\mu$m)  &whole SNR &E blob &source\\
\hline
3.2 &  (3.6 $\pm$ 0.4)$\times$10$^{-2}$&$\cdots$&(1)\\
3.6 &  (3.8 $\pm$ 0.6)$\times$10$^{-2}$&(6.0 $\pm$ 0.8)$\times$10$^{-3}$&(2)\\
4.5 &  (3.2 $\pm$ 0.5)$\times$10$^{-2}$&(7.4 $\pm$ 0.8)$\times$10$^{-3}$&(2)\\
5.8 &  (1.8 $\pm$ 0.9)$\times$10$^{-1}$&(3.0 $\pm$ 0.4)$\times$10$^{-2}$&(2)\\
7.0 &  (2.8 $\pm$ 0.3)$\times$10$^{-1}$&$\cdots$&(1)\\
8.0 &  (2.3 $\pm$ 1.0)$\times$10$^{-1}$&(4.1 $\pm$ 0.5)$\times$10$^{-2}$&(2)\\
11  &  (3.3 $\pm$ 0.3)$\times$10$^{-1}$&$\cdots$&(1)\\
15  &  (8.7 $\pm$ 0.9)$\times$10$^{-1}$&$\cdots$&(1)\\
24  &  1.6 $\pm$ 0.1&(30 $\pm$ 0.6)$\times$10$^{-2}$ &(2)\\ 
70  &  9.9 $\pm$ 0.5&(18 $\pm$ 0.9)$\times$10$^{-1}$ &(2)\\
100 &  8.5 $\pm$ 1.7&1.4 $\pm$ 0.4&(3)\\
160 &  7.2 $\pm$ 1.4&1.2 $\pm$ 0.2&(3)\\
250 &  2.1 $\pm$ 0.3 &(5.3 $\pm$ 0.8)$\times$10$^{-1}$&(3)\\
350 &  (5.0 $\pm$ 0.8)$\times$10$^{-1}$  &(2.8 $\pm$ 0.4)$\times$10$^{-1}$ &(3)\\
500 &  (1.0 $\pm$ 0.2)$\times$10$^{-1}$ &$\cdots$ &(3)\\
\hline
\end{tabular}
\end{table}

\begin{figure}
\centering
\includegraphics[width=7.0cm]{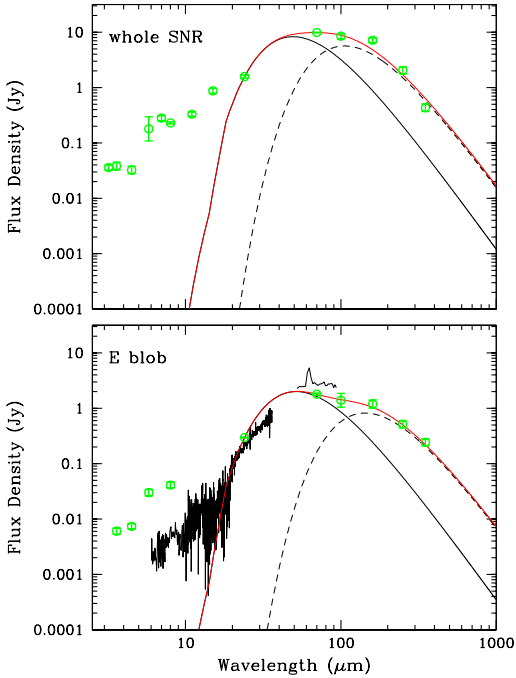}
\caption{
SED plots of the whole SNR ({\it upper panel}) and 
E blob only ({\it lower panel}) together with silicate dust fits.
The strong emission lines of atomic gas and H$_{2}$ in the IRS spectrum
 are subtracted. The black solid and broken lines are the SEDs of warm
 and cold components, reflectively. 
The red lines are the sum of these two components.
\label{model}}

\end{figure}

\begin{table}[t] 
\centering
\caption{The derived dust mass and temperature.}
\begin{tabular}{@{}lccccc@{}}
\hline
part  &dust  &comp.&$T_{d}$   &$M_{d}$\\
      &spices&&(K)       &($M_{\odot}$)\\
\hline
SNR   &carbon  &warm &62 $\pm$ 2  &0.10 $\pm$ 0.02\\
      &         &cold &26 $\pm$ 2  &8.7 $\pm$ 1.0\\
      &silicate &warm &57 $\pm$ 2  &0.4 $\pm$ 0.06\\
      &         &cold &27 $\pm$ 2  &12 $\pm$ 2\\
\hline
E blob&carbon   &warm&59 $\pm$ 2&0.04 $\pm$ 0.01\\ 
      &         &cold&19 $\pm$ 2&4.2 $\pm$ 1.7 \\ 
      &silicate &warm&57 $\pm$ 1&0.10 $\pm$ 0.01\\ 
      &         &cold&21 $\pm$ 1&6.1 $\pm$ 2.2\\
\hline
\end{tabular}
\end{table}

\section{Dust mass in the N49 system and discussion}
We estimated the dust mass with the line-contribution-subtracted data
and its temperature through SED fitting. First, we estimated the 
contribution of the synchrotron emission to the mid- to far-IR region.
Dickel \& Milne (1998) measured the flux densities at 3.5/6/13/20 cm 
for the whole SNR. By extrapolating the flux densities at this radio
wavelength to the mid-IR region, we found that the 
contribution is negligibly small ($<$2 $\%$) to the 24/70/100/160/250
$\mu$m flux densities. In the SPIRE 350/500 $\mu$m bands, the
contribution is large; $\sim$14 $\%$ for 350 $\mu$m and
$\sim$79 $\%$ for 500 $\mu$m. Before fitting, we subtracted the contribution 
of the synchrotron emission to the 350 $\mu$m flux density for the whole
SNR and E blob, assuming that the contribution can be also applied 
to the E blob. In the fittings we excluded 500 $\mu$m data.

When we assume optically thin thermal radiation at a dust
temperature 
$T_{d,i}$ of a component $i$ and a single dust species with the 
radius $a$, the observed 
flux density $F_{\nu}$ is written by

\begin{equation}
F_{\nu} = \left(\frac{4}{3}a\rho D^{2}\right)^{-1} \times \sum_{i} 
M_{d,i}B_{\nu}(T_{d,i})Q_{\nu,i},
\label{dust}
\end{equation}

\noindent 
where $M_{d,i}$ is the dust mass of the component $i$, $B_{\nu}$($T_{d,i}$) is the Planck function of $T_{d,i}$, and 
$Q_{\nu,i}$ is the absorption efficiency, $\rho$ is the dust grain density and 
$D$ is a distance to the observed dust from us (Kwok 2007). Here, we adopted $D$=50 
kpc (Schaefer 2008).  We considered two types of grains, amorphous carbon and silicates.  For amorphous carbon the
optical constants were taken from Zubko et al. (1996). We adopted
$\rho$=2.26 g cm$^{-3}$. To simplify the model, we assume $a$ = 0.1$\mu$m uniform grain radius. 
Next, we considered astronomical silicate only. The optical constants were taken from Draine 
\& Lee (1984) and a $\rho$ of 3.6 g cm$^{-3}$ was adopted. Due to the
lower absorption efficiency at $>$70 $\mu$m than amorphous carbon, a larger dust
mass would be estimated. The dust mass estimated for the silicate-only case would be an upper limit.
In the fittings we found the best fit $M_{d,i}$ and $T_{d,i}$ to
the observations by least-square technique. In this process, 
$M_{d,i}$ and $T_{d,i}$ are the free parameters, and all
the data are equal of weight.

We first considered a single temperature model to fit the SED. For the whole
SNR, the best fit had a $T_{d}$ of 33 K and 34 K for amorphous
carbon and for silicate respectively, and a derived $M_{d}$ of 3.6 
$M_{\odot}$ and 4.6 $M_{\odot}$. For the E blob, $T_{d}$ is 35 K (carbon case) and 
39 K (silicate case), and $M_{d}$ is 0.47 $M_{\odot}$ and 0.60 $M_{\odot}$. 
The temperature is fairly similar to the temperature derived by 
Graham et al.\ (1987) from {\it IRAS} data. However, we found that a better fit could be obtained
if we considered a model with two-temperatures instead of one. The 
one-component fittings cannot simultaneously explain the $<$24 $\mu$m 
and $>$160 $\mu$m flux. The implied dust masses and temperatures for the two-temperate model are 
summarized in Table 2 for the two types of grains.

The best-fit two-temperature model, assuming silicate grains, for the
SEDs is shown in Fig 2.  The warm component has a 
temperature of $\sim$60 K, and is surely due to dust in shocked gas,
while the cold component has a temperature
of $\sim$20 K, similar to that found in molecular clouds (16-24 K,
Bernard et al. 2008; Gordon et al. in 2010). The contribution from the cold
component to the total flux in the {\it Herschel} bands is $>$80 $\%$. 
The dust mass and temperature of the  warm component in the E blob are 
very close to van Loon et al. (2010; 0.2 $M_{\odot}$ and 43 K).
The total mass of dust associated with N49 is about 10 $M_{\odot}$ and is dominated by dust at cold temperatures, dust which was difficult to
observe with {\it Spitzer}.
It is tempting therefore to identify the warm component with dust in the
X-ray plasma and the cool component not
with the SNR, but with the molecular cloud.  However, this assignment is
very uncertain, because dust temperature variations are also expected 
as a result of the large variations of post-shock gas density that are known to exist in
N49.

To place SNR N49's dust mass estimate into the context
of SNR development, we plot the dust mass vs. age for all LMC and SMC
SNRs studied to data, as noted in Fig 3. Note that except for N49 the dust masses are
estimated based on $<$70 $\mu$m data. The relation can be with the free 
expansion phase early on, followed by a Sedov phase (radius $\propto$
(age)$^{2/5}$; cf. Badenes et al. 2010).

The mass of the warm dust component in N49 is 0.1 $M_{\odot}$ assuming
amorphous carbon and 0.4 $M_{\odot}$  assuming silicates.  Adopting a dust-to-gas mass 
ratio of $\sim$250 (Meixner et al. 2010) and assuming that 
only the warm dust is associated with shocked gas, the swept-up mass 
of the SNR ranges from 25 to 100 $M_{\odot}$.  
This is comparable to the mass estimated from X-ray 
observations, 200
$M_{\odot}$ (Hughes et al. 1998), assuming the true grain characteristics are similar to those of
silicates.  
A much higher  (25$\times$ larger)  swept-up mass is estimated if we included
the cold dust in our estimate, which suggests that in the context 
of our simple model there is a significant molecular
cloud component to the cold dust we associate with the SNR, and 
that the shocks propagating into the molecular cloud may be too slow 
to produce X-rays.  While more sophisticated modeling of N49 may refine
the mass estimates somewhat, the basic limitation in arriving at a 
more precise picture of the nature of the interaction with the molecular 
cloud and the amount of shocked molecular gas in N49 is the limited
angular 
resolution of {\it Herschel} at longer wavelengths.

\begin{figure}
\centering
\includegraphics[width=7.0cm]{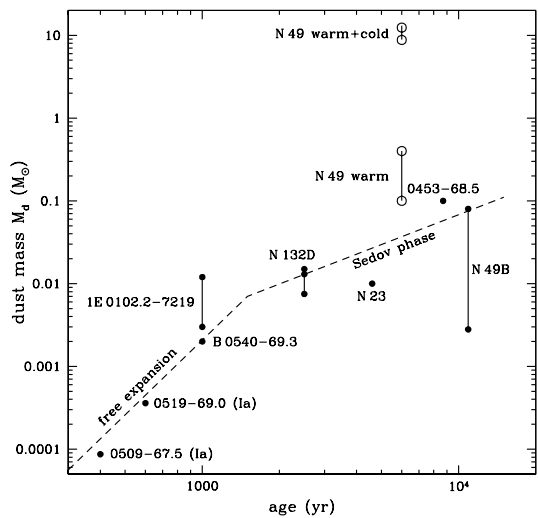}
\caption{The relation between the dust mass and age. The data are taken
 from Seok et al. (2008) for 0509-67.5, 0519-69.0, N132D, and N49B; 
Williams, B. et al. (2008); Sandstrom et
 al. (2009) and Rho et al. (2009) for 1E 0102.2-7219; Williams, B. et
 al. (2006) for N132D, N23, 0453-68.5, and N49B; van Loon et
 al. (2010) and present work for N49.\label{model}}

\end{figure}

\section*{Acknowledgments}
We acknowledge financial support from the NASA {\it Herschel} Science Center, 
JPL contract \# 1381522.  We are grateful for the contributions and support from 
the European Space Agency (ESA), the PACS and SPIRE teams, 
the {\it Herschel} Science Center and the NASA {\it Herschel} Science Center 
(esp. A. Barbar and K. Xu) and the PACS and SPIRE instrument control 
center at CEA-Saclay,  without which none of this work would be possible.

\end{document}